\documentstyle[eqsecnum,aps,psfig]{revtex}
\draft

\begin{document}

\title{Relaxation fluctuations about an  equilibrium in quantum chaos}
\date{\today}
\author{Arul Lakshminarayan \cite{byline} }
\address{ Physical Research Laboratory,
Navarangapura, Ahmedabad,  380009, India.}
\maketitle

\begin{abstract}
Classically chaotic systems relax to coarse grained states of
equilibrium.  Here we numerically study the quantization of such
bounded relaxing systems, in particular the quasi-periodic
fluctuations associated with the correlation between two density
operators. We find that when the operators, or their Wigner-Weyl
transforms, have obvious classical limits that can be interpreted as
piecewise continuous functions on phase space, the variance of the
fluctuations can distinguish classically chaotic and regular motions,
thus providing a novel diagnostic devise of quantum chaos. We uncover
several features of the relaxation fluctuations that are shared by
disparate systems thus establishing restricted universality.  If we
consider the nonlinearity driving the chaos as pseudo-time, we find
that the onset of classical chaos is indicated quantally as the
relaxation of the relaxation fluctuations to a Gaussian distribution.

\end{abstract}

\section{Introduction}

\hspace{.5in} Classical dynamical systems can be classified into a
hierarchy of deterministic randomness. We have well studied examples
from integrable systems to Bernoulli systems, from the regular to
those that are in a coarse grained manner identical to stochastic
processes.  An important notion in this context is that of mixing, as
exemplified in the now classic coke-rum mixture of Arnold and
Avez \cite{ArnAvez}. Apart from the fundamental role it 
plays in deterministic
randomness, alias chaos, it is the backbone of the foundations of
classical statistical mechanics. While ergodicity is compatible with
equilibrium, it is mixing that would ensure the drive to this state.

Quantum dynamical systems are not so neatly categorized. Quantum
mixing remains a difficult notion. There are specific examples where
indeed mixing can be said to occur in the configuration space, in as
much as any two spatial wavefunctions decay and decorrelate exactly as
classical Liouville phase space densities \cite{Weigert}. Yet these are rather
special systems generally open and with a continuous spectra. They do not
address the issue in more generic situations. The phenomenon of quantum 
suppression of classical chaos in diffusive systems is now well studied,
and several localization mechanisms have been put forward which inhibit
quantum mixing \cite{Izrailev}, yet the issues in the situation of ``hard chaos'' for 
bounded chaotic systems relaxing to an equilibrium have not been sufficiently
addressed. 
. 

This paper studies quantum objects that are obviously connected to the twin
issues of ergodicity and mixing in the classical limit. In the process 
we use models having a discrete spectra whose classical limit is known to
be chaotic. The most convenient for our purposes is the quantization 
of two dimensional area preserving maps on the torus- a much studied subject.
The Hilbert space is then finite dimensional and we have fully all the 
contradictions between quantum and classical chaos, while retaining the 
attractive feature of being able to do the quantum mechanics up to machine
precision and size. 
 
Let the relevant classical phase space be $\Omega$. This could be for
instance the energy shell on which the Hamiltonian flow is restricted.
We may formulate mixing either in terms of phase space functions or
densities. Let $A$ and $B$ be two subsets of the phase space such that
they do not intersect. Let $\chi_{A}(x)$ and $\chi_{B}(x)$ be the two
characteristic functions of these subregions. Let the flow integrated
for a time $t$ be denoted by $f^{t}$, this could be a continuous flow
or a discrete map, we formulate this for the case of maps in which
case $t$ is an integer. The invariant measure of the flow is denoted by
$\mu$, in the present context of area preserving maps this is simply the 
physical area of the phase space region. 

The central quantity of interest is the correlation between the
functions $\chi_{A}$ and $\chi_{B}$. In more graphic language, as the
subregion $A$ evolves with the motion the correlation is the
fractional area of its intersection with the region $B$. If in the
long time limit this factorises into the fraction of the areas of $A$
and $B$ we have mixing. In other words the fraction of $A$ systems in $B$
is the fractional area of $B$ in the long time limit. 
\begin{equation}
\mu(f^{t}(A) \cap B)/\mu(A) \longrightarrow  \mu(B)/\mu(\Omega)
\end{equation}
Mixing systems are a step above ergodic ones
in the hierarchy of classical dynamical systems. Ergodicity is the equality
of the time average and phase space average of almost all points in the 
phase space. This can also be formulated as decorrelation on the average.
Thus a system is ergodic if
\begin{equation}
 \lim_{T \rightarrow \infty} \frac{1}{T} \sum_{t=1}^{T} \mu(f^{t}(A) \cap B)/\mu(A) \, =\,  \mu(B)/\mu(\Omega).
\end{equation} 

Classical mixing systems are characterized by a series of complex numbers
or resonances that dictate the rate of decay of correlations. Thus for 
purely hyperbolic or Axiom A systems we can write
\begin{equation}
\mu(f^{t}(A) \cap B)/\mu(A)-  \mu(B)/\mu(\Omega)=\sum_{i} C_{i} \exp(\lambda_{i}t). 
\end{equation}
The $\lambda_{i}$ are the so called Ruelle resonances and are in
general complex numbers with negative real parts independent of the
particular partitions $A$ and $B$. Explicit 
calculation of these
resonances is a challenging part of dynamical systems theory; there
are special models which are isomorphic to finite Markov processes for
which these can be analytically found; for example the multi-baker
maps \cite{ElsKap,GaspJsp92}.

What we wish to study below are quantities in quantum mechanics that
are naturally related to the classical correlation functions defined
above. In particular we will analyze the long time behaviour of these
functions as we expect there to be purely quantum effects in these
regimes.  For short times it is expected, and has been demonstrated,
that the classical and quantum largely coincide \cite{UBT,LBJsp}. 
The long time
behaviour is of course very interesting because equilibrium is an
asymptotic (in time) property. Just as there is an ``equilibrium''
classically, a state where complete decorrelation has been practically
achieved, we will study the quantum equilibrium and the associated
fluctuations which we will find has universal features in a limited manner 
and is often a
sensitive measure of whether the classical limit is chaotic or not. The 
qualifications in the previous statement are necessary for reasons we will 
clarify below.

\section{The Quantum Correlation Function}

The relevant dynamical space on quantization is a Hilbert space
$H_{N}$. For regions of phase space we consider subspaces of the
Hilbert space. Since we have in mind classical maps on the two torus,
let the Hilbert space be of finite dimensionality $N$.  Let there be
two distinct density operators on $H_{N}$ (may or may not be projection
operators), $P_{A}$ and $P_{B}$.  We will study the cases when these
operators on $H_{N}$ have obvious classical limits as functions on
the phase space.  We will restrict ourselves initially, and largely,
to the case when the operators are projection operators such that
$P_{A}+P_{B}=I_{N}$, where $I_{N}$ is the $N$ dimensional identity
matrix. Classically this corresponds to choosing the partitions $A$
and $B$ such that $A \cup B=\Omega$.

We sidestep the question of assigning an operator to a general
subregion of phase space, and whether this is possible at all (say
through the Wigner-Weyl transform of the characteristic function), by
restricting this study to particularly simple subspaces (or density operators)
both of $\Omega$ and on $H_{N}$ and going from the quantum to the
classical instead of vice versa.  Choosing an ``arbitrary '' subspace
( or operator) of $H_{N}$ obviously does not correspond to a proper
subregion of the classical phase space and we will see that such
subspaces do not provide interesting relaxation behaviors in that
they do not distinguish between classically chaotic and regular
motions. We may also construct operators out of coherent states that
provide an adequate quantum equivalent of the characteristic
functions.  For instance If $|z \rangle $ is the usual coherent state
on the plane, we may consider
\begin{equation}
P_{A}= \int_{A} d^{2}z \, |z \rangle \, \langle z|.
\end{equation}
We will finally use such operators after adapting them to the 
toral phase space. 
  
If we use pure states, say  $|n \rangle $, we can  construct the 
density operators as $\sum |n \rangle \langle n  | $.
Let the dimensionality of the subspace $A$ be $n_{A}=f_{A}N$; then
$n_{B}= N-n_{A}=f_{B}N$. The fractions $f_{A}$ and $f_{B}$ determine
the relative sizes of the partitions and we will assume below that
$1<<n_{A},n_{B}<<N$, that is the subspaces are large enough to contain
very many states and be considered the quantum equivalent of a finite
(much larger than $\hbar$) region of phase space. We will mostly use
projection operators that are diagonal in the position basis. In this
case the classical region corresponding to 
\begin{equation} P_{A}= \sum_{n=0}^{f_{A}N-1} |n \rangle \langle n|
\label{proj-op}
\end{equation} 
is the rectangle
$[0,f_{A}) \times [0,1)$ and the region corresponding to $P_{B}$ is
the rectangle $[f_{A},1) \times [0,1)$, as the phase space is taken to
be the unit torus $(q,p)=[0,1) \times [0,1)$.

We now introduce the dynamics as an unitary operator $U$ acting on the
states of $H_{N}$. The central quantity of interest is then
\begin{equation}
C_{AB}(t)=\mbox{Tr}(U^{t}P_{A}U^{-t}P_{B})/N,
\end{equation}
 which gives us the overlap of the subspace $A$ propagated for time
$t$ with the subspace $B$. We may note that this is an important and
natural physical quantity to study and has appeared before in several
contexts, for example ref. \cite{UBT}. It can also be viewed as an 
analogue of Landauer conductance
for bound systems and as such should be of considerable importance in
quantum transport.  If in fact there is factorization in time then
$C_{AB}(\infty)$ must be compared with
$\mbox{Tr}(P_{A})\mbox{Tr}(P_{B})/N^{2} = f_{A}f_{B}$. The operators
$U$ used in this paper are briefly described in the appendix and have 
been essentially described before.

Since $U$ has a discrete and finite spectrum, the correlation can only
be a finite sum of purely oscillatory terms and can therefore display
decay only over short time scales (of the order of Heisenberg
time). The decorrelation is therefore not expected. As an extreme
quantum example if we consider the case when the projection operators
are constructed out of the basis functions of $U$ the correlation is
zero for all time. This is an extreme case and for instance it is not
clear what the classical limit is of the Wigner-Weyl transform of 
projection operators
constructed out of such a basis.  We will consider ``generic''
subspaces and projection operators and any claim of universality made
below has to be viewed with this caveat. We may note that this plagues
Random matrix theory descriptions of eigenfunctions as well, as it
involves basis dependent quantities. Here the basis dependence enters
as the basis in which the projection operators are diagonal.

It was suggested earlier using quantized multi-baker maps that the
quantum correlation function approached the classical correlation
function in the classical limit corresponding here to $N \rightarrow
\infty$ \cite{LBJsp}.  Certain remarkable features of quantum 
relaxation were noted
there including relaxation localization and effects of symmetries on
transport. Here we wish to study the fluctuation properties more
closely and uncover possible universal features. Therefore we study
rather standard models like the Taylor-Chirikov (standard) map \cite{Izrailev}, the
kicked Harper system \cite{Leb} and the baker map \cite{BVbak} 
although most of these models
do not have explicit expressions for the classical Ruelle resonances
and are probably not Axiom A systems (except the baker map). However
it is very reasonable to assume that in certain well known parameter
regimes they can be for all practical purposes mixing systems.

Normalizing the correlation so that the classical limit, if it exists,
is for large times unity we study the quantity
$c(t)=C_{AB}(t)/f_{A}f_{B}$, where from now we will acknowledge
implicitly the dependence of the correlation on the particular choice
of partitions. Fig. 1   shows an example of the generic behaviour of
$c(t)$, this particular data being for the standard map at two
different values of the inverse Planck constant $N$ for the same
classical value of the chaos parameter. The initial relaxing behaviour
is not clear from the figures as the time scales are much larger than
the inverse of the principal Ruelle resonance. The projection
operators used are diagonal in the discrete position basis, $P_{A}$ is
given by Eq.( \ref{proj-op})
and $P_{B}=I_{N}-P_{A}$. The position eigenbasis is denoted as $|n \rangle $
and in this figure $f_{A}=f_{B}=1/2$. 

We note that in this paper we have 
used dimensionless scaled position and momentum co-ordinates and that the 
time is measured as multiples of the period of the kick, taken as unity. 
Modulo one conditions restrict the phase space to the unit torus so that 
the dimension of the Hilbert space $N$ is related to the Planck constant as 
$N=1/h$ and the classical limit corresponds to $N \rightarrow \infty$
while the parameters of the map such as the  kick strengths 
are scaled and dimensionless.

The first observation is that the quantum correlation is in fact
quite close to the classical value of unity, second is that the
average of the oscillations in relation to unity will give us some
information about whether quantum mechanics is inhibiting transport or
otherwise, thirdly this average must in the classical limit tend to
unity, fourth is the observation that the fluctuations are getting
smaller in the classical limit and must tend to zero. Some of these
observations have been made and substantiated earlier \cite{LBJsp}, here we will
elaborate and present more results. 

Previous work on the use of time developing states to study the quantum 
manifestations of chaos have basically concentrated on the survival probability
of a pure state \cite{SurvProb}. This corresponds to the choice 
$P_{A}=P_{B}=|\psi \rangle
\langle \psi |$ with an arbitrary normalized initial state $|\psi \rangle$ and thus
$f_{A}=f_{B}=1/N$. Such survival probabilities averaged over random initial states show a 
distinct difference between chaotic and regular systems. In contrast what we study in 
this paper are density (or projection) operators and not pure states. It becomes quite 
important that $f_{A}N$ and $f_{B}N$ are large and are of the order of the Hilbert space
dimensionality $N$ itself. Besides any ``arbitrary'' state can be chosen for the 
calculation of the survival probability while we will restrict ourselves to those 
projection operators that can be interpreted as phase space regions in the classical limit.
This is to ensure that the classical  transition to chaos is fully reflected in the quantum 
relaxation, as will be illustrated below.

\subsection{The average}

The quantum ``equilibrium'' as opposed to the classical is an highly
oscillatory state, with the fluctuations coming from the discrete
nature of the spectrum. We will denote the time average of the
correlation function by $<c>$ and its variance by $\sigma^{2}$. These
quantities can be written in terms of the eigenfunctions of the
evolution operator $U$, which satisfy the following equation.
\begin{equation} 
 	U|\psi_{m} \rangle = \mbox{e}^{i E_{m}} |\psi_{m} \rangle \; \; 
m=0,\ldots N-1
\end{equation}
The real numbers $E_{m}$ are the eigenangles of the quantum map, and we 
are assuming that there are no exact degenerecies as is the generic case.  
The correlation is then
\begin{equation}
c(t)=\frac{1}{N f_{A} (1-f_{A})} \sum_{m_{1},m_{2}} \sum_{n_{1}=0}^{
Nf_{A}-1} \sum_{n_{2}=Nf_{A}}^{N-1} \mbox{e}^{i t (E_{m_{2}}-E_{m_{1}})} \\
\langle n_{2}|\psi_{m_{1}} \rangle \langle \psi_{m_{1}}|n_{1} \rangle 
\langle n_{1} |\psi_{m_{2}} \rangle \langle \psi_{m_{2}}|n_{2} \rangle 
\end{equation}
Thus we can express the average quite simply as 
\begin{equation}
<c>=\frac{1}{N f_{A}(1-f_{A})} \sum_{m=0}^{N-1} \sum_{n=0}^{f_{A}N-1}
\sum_{n^{\prime}= f_{A}N}^{N-1} |\langle n|\psi_{m} \rangle |^{2}
|\langle n^{\prime}|\psi_{m} \rangle |^{2}.
\end{equation}

Thus the average is a measure of the distribution of the eigenvectors
in the subspaces $A$ and $B$. We suggest that $<c> \, < \, 1$
generically, indicating a certain reluctance to participate equally in
both the partitions in proportion to their sizes.  Time dependent
quantum chaotic systems such as the kicked rotor on the cylinder are
known to suppress classical chaos and lower diffusion \cite{Izrailev}. The
corresponding suppression in the case of bounded systems is in the
average relaxation such as measured by $<c>$.  
We note that the sum over $n$ and $n^{\prime}$
expressing the average can be factored into two single sums, and we
see that if for instance we had a parity symmetry forcing the
wavefunction to be essentially identical in the subspaces $A$ and $B$
we would have $<c>=1$, which is indeed the case for the fluctuations
shown in Fig. 1.  The average $<c>$ could also be greater than unity
in the presence of symmetries, or if the partition $B$ is identical to
$A$. Although in the absence of any special symmetry it is true that
classical chaos will tend to enhance the average making it closer to
unity, this effect is quite marginal and practically non-existent when 
a transition to complete classical chaos is achieved. 

Fig. 2a  shows the average for the quantized baker map implementing the
Bernoulli scheme $(2/3,1/3)$ as a function of the inverse Planck
constant $N$. The partitions are such that $f_{A}=f_{B}=1/2$. Shown is
the deviation of the average from unity and an approximate power law
is observed for large $N$. Thus we can write
\begin{equation}
<c> \, \sim \, 1-\alpha N^{-\gamma},
\end{equation}  
and in this case $\gamma \approx 3/4$. The relaxation localization is
implied by $\alpha$ being positive.  
Fig. 2b shows the same quantity for the standard map and we find 
for the partition $f_{A}=1/4, f_{B}=3/4$ that $\gamma \approx 1$,
with a different value of $\alpha$. We note that in both the cases there
is a small oscillation about the fitted straight line.
Fig. 3a shows the variation of the
average with the classical kick parameter $K$ for the standard map, the
map undergoing a transition to complete chaos with the breaking of the
last KAM torus at $K \approx 1$. The partition $A$ was such that
$f_{A}=1/4$ and $B$ was the complimentary space. Also illustrated is
the localization effect and the same for the kicked Harper model
in Fig. 3b. The kicked Harper map, or simply the Harper map, undergoes
a transition to chaos at the kick strength $g \approx 0.63$ (see Appendix).

\subsection{The variance and the distribution}

A promising candidate is the variance of the fluctuations which
exist irrespective of the symmetries of the system. This can also be 
expressed in terms of the eigenfunctions of the system. In the particular 
case when $P_{A}+P_{B}=I_{N}$, writing for $f_{A}$ simply $f$,
we get after some simplifications 
\begin{equation}
\sigma^{2}= \frac{2}{N^{2} f^{2}(1-f)^{2}} 
\sum_{m_{2}>m_{1}} \left | \sum_{n=0}^{fN-1} \langle n|
\psi_{m_{2}} \rangle \langle \psi_{m_{1}}|n \rangle \right|^{4}.
\end{equation}
Thus the variance measures a correlation between distinct 
pairs of eigenfunctions. There is a coherent partial sum over $Nf$ 
states that makes the variance non-trivial and perhaps non-universal. 

Fig. 4a shows the standard deviation $\sigma$ scaled to $\sigma
N$ as a function of the kick strength in the case of the
standard map. The transition to classical chaos at $K \approx 1$ is
visible in the variance of the fluctuations as a point at which it
attains an approximately constant value. The relaxation fluctuations
are therefore significantly different depending on the dynamical
nature of the classical limit, being in general larger for regular
systems than chaotic ones, and thus provide a novel diagnostic devise
for the study of quantized chaotic systems.

Fig. 4b shows identical quantities for the case of the kicked Harper
map. 
The variance seems to have no further dependence on measures of chaos
and scales as $\hbar$ quite universally when the classical limit is
chaotic.  This scaling is shown explicitly in Fig. 5 for three
systems in the chaotic regime: the kicked rotor, the kicked Harper
 map and the usual quantum
bakers map. The fact that the three lines, of slope close to $-1$, are
practically on top of each other is evidence of a limited universality
of the relaxation fluctuations. The limits to universality will be
discussed in the following, but also shown in the figure are the
results for a generalized bakers map implementing the $(2/3,1/3)$
Bernoulli scheme which clearly shares the same slope but is different
otherwise from the other cases which are symmetric. 

When the classical limit is regular there is still evidence for
scaling with $\hbar$ but with non-universal exponents. Thus Fig. 6
suggests that for the standard map, 
\begin{equation}
\sigma \sim N^{-\gamma},
\end{equation}
with the exponent $\gamma$ increasing to unity as the kick strength is
increased to induce the transition to chaos. When $K=20$ or $K=30$
the exponent is already close to unity and the two cases 
are practically indistinguishable.

When $f \ne 1/2$ the relaxation is between non-symmetrically related
partitions of unequal measures. The quantum variance is studied 
directly as a function of $f$. Still we are considering the case when 
the two partitions complement each other. Results not shown here 
indicate that the dependence  of the standard deviation on $N$ (or $\hbar$)
remains the same, namely $N^{-1}$. Thus we may conjecture the existence of a 
function $g(f)$ such that $\sigma\, =\, g(f)/N$. The function $g(f)$ will 
naturally depend on the choice of the density operators $P_{A}$ and $P_{B}$. 
If further we choose these operators as the projection operators 
described by the Eq.( \ref{proj-op}), we numerically find the relation
\begin{equation}
\sigma \, =\, \frac{\pi^{2} (2 f)^{\beta}}{16 N f (1-f)} \; \; 0<f\le 1/2
\end{equation}
where $\beta \approx 0.83$. The constant has been identified as related 
to the number $\pi$ by averaging over different $K$ in the chaotic regime. 
Beyond $f=1/2$ symmetry of the maps ensures that $\sigma(f)=\sigma(1-f)$.
It is to be emphasized that although $\sigma$ is independent of model 
parameters like $K$, we are assuming that there has been a completed 
transition to full scale chaos. 

Fig. 7 compares $\sigma N f(1-f)$ with the fit $\pi^2 (2f)^{\beta}/16$
for the standard map, the Harper map and the bakers map. The standard map
and the Harper fall practically on top of each other while the bakers map
shown deviations although the trend is maintained. The insensitivity of this
curve to the parameter $K$ or $g$ indicates that once chaos is achieved 
the fluctuations are essentially the same. The curve $\sigma N$ as a function
of $f$ is thus interesting in that it has restricted universal features. 
The Harper and the standard maps behave almost identically while there 
are deviations for the baker; moreover the curve is most definitely dependent 
on the operators $P_{A}$ and $P_{B}$, but once this is fixed the fluctuations 
might be model independent.

Though the fluctuations are not universal they do distinguish
classically regular and chaotic motions when the operators $P_{A}$ and
$P_{B}$ have meaningful classical limits. If this condition is
violated then the fluctuations are incapable of such a
separation. This is in contrast to RMT based arguments for the
randomness of a wavefunction, where so much information is lost that
in any generic basis the wavefunctions of a classically chaotic system
are universally distributed. We will illustrate this in two ways. In
the first we choose $P_{A}=\sum_{ n \; \mbox {even }}|n \rangle \langle n
|$ and $P_{B}$ is the complementary space. Clearly in the classical
limit these projection operators do not describe meaningful phase
space regions. In the second we choose half the eigenbasis of the
kicked rotor at $K=155$ for constructing $P_{A}$ and $P_{B}$ is the
complementary space. 
Fig. 8 shows that the variance does not capture the transition to
chaos at around $K=1$ for the kicked rotor, unlike in the cases above,
one of which has been repeated for comparison purposes.  However once
the transition to chaos is complete the three standard deviations seem 
to coincide. We have diagonalized a standard map at $K=155$ to 
ensure that there are no special correlations with the system at around 
$K=4$. 

Finally we construct the operators out of coherent states so that the
interpretation as subspaces of phase space is intuitively obvious in
the classical limit. We use the discrete toral coherent states
developed by Saraceno \cite{Sarabak}. Let $|m,n \rangle $ be such a 
state that is
localized at $(q=m/N, p=n/N)$ on the unit torus. We take 
\[
P_{A}= \sum_{(q,p) \in  A} |m,n \rangle \langle m,n |
\]
and $P_{B}$ is similarly constructed. In the computation shown below
we have taken $A$ to be the subspace $[0,1/2) \times [0,1/2)$ and $B$
to be $[1/2,1) \times [1/2,1)$. The variance is evaluated as above
with some minor differences, since $P_{B} \ne 1-P_{A}$.  The measures
of the quantum subspaces are $N^{2}/4$, as we have used normalized
coherent states. The eigenfunctions are expressed in the coherent
state basis and the variance is evaluated.  Fig. 9 shows the results
of this computation and once more we recover the property that the
variance captures the classical transition to chaos.
Not presented are some more results, in particular the computations using the 
momentum basis vectors, as these simply reflect the general trends already 
noted above.

We have been studying the mean and standard deviation of a quantity that 
is truly a quasi-periodic oscillation;
the distribution of such a quantity 
ought to be significantly different from a random process. However 
Fig. 10 motivates that the effect of quantum chaos is to cast the oscillations 
into the ubiquitous Gaussian process, and increasingly more so in the 
classical limit, thereby indicating that the mean and standard deviation 
are sufficient to characterize the relaxation process. The solid line is 
a Gaussian distribution with zero mean and standard deviation equal to 
$\pi^{2}/4$, as the partitions used correspond to  $f_{A}=f_{B}=1/2$. 
In sharp contrast Fig. 11 shows the same for the case of the standard map
at $K=0.2$ when the dynamics is predominantly regular. The distribution is 
much broader and is also non-symmetric. Thus if the parameter $K$ is considered
as a pseudo time variable, the onset of chaos is signaled as the realization
of a stationary state of the relaxation fluctuations - the Gaussian 
distribution.

\section{Summary}

 We have studied the quasi-periodic relaxation in bound quantum systems
whose classical counterparts relax to a coarse grained equilibrium. We
have concentrated on the correlations between density operators whose
classical limits (or more correctly the limit of their Wigner-Weyl
representations) may be interpreted as piecewise continuous functions on
phase space. We have observed that in the absence of special symmetries the
average quantum relaxation is smaller than the classical, and we have
presented some scaling relations with $\hbar$. 

The relaxation fluctuations have been shown to provide rather
interesting characterization of the quantum motion. They sometimes
have universal characteristics although these are limited. The kicked
map and the Harper model with identical symmetry properties have been
shown to be practically identical in the behaviour of their relaxation
fluctuations. The standard deviation of these fluctuations have been
studied and has been shown to decrease to a constant value with a
rather sharp change in the values corresponding to the transition to
chaos in the classical system. Thus the fluctuations provide a new way
to characterize quantum chaos. The standard deviation has been shown
to scale as $\hbar$ when there is classical chaos universally, while
it has other non-universal power law dependencies when the classical
motion is not chaotic.

Random matrix theories have been providing a framework to 
study several properties of quantized chaotic systems and 
it is natural to explore the random matrix predictions for the 
quantities studied in this note. It is of special interest to place 
the numerical results in this paper in the context of a generalization 
of the previous work on survival probabilties \cite{SurvProb} and 
work on these aspects  are 
currently underway. 

I thank an anonymous referee for pointing  out the references in \cite{SurvProb}
as well as for some useful comments.

\appendix 
\section*{}

The Taylor-Chrikov, or the standard map, or the kicked rotor, used in
this paper is described below \cite{Izrailev,ShiChang}.  
Let the classical map be
$(q_{t+1}=q_{t}+p_{t+1}, \; \; p_{t+1}=p_{t}-V^{\prime}(q_{t+1}))$,
both $q $ and $p$ taken mod 1.
$V^{\prime}$ is the derivative of the kicking potential which is
assumed to have unit periodicity. The toral states are assumed to
satisfy certain boundary conditions specified by a point on the dual
torus.  Let $|q_{n} \rangle $ and $|p_{m} \rangle $ be the position
and momentum states then $|p_{m+N} \rangle =e^{-2 \pi i a } |p_{m}
\rangle $ and $|q_{n+N} \rangle =e^{2 \pi i b } |q_{n} \rangle $, where
$(a,b)$ are real numbers between 0 and 1;  $N$ is the dimensionality
of the Hilbert space. If $b=0$, upon canonical quantization we get the finite unitary
operator 
\begin{equation} 
U_{n, n^{\prime}}= \frac{e^{i \pi/4}}{\sqrt{N}} \exp (-2 \pi i N
V(\frac{n+a}{N})) \exp(i \frac{\pi}{N}(n-n^{\prime})^{2}).
\end{equation}
In this paper we have used for the standard map $V(q)=K \cos(2 \pi
q)/(2 \pi)$, and $a=1/2$ which makes the quantum map possess an exact
parity symmetry about $q=1/2$. The values of $N$ are restricted to the 
even integers.

The kicked Harper map \cite{Leb} on the torus is similar to the above system
except for the momentum dependence. If the Hamiltonian is
\[ H=V_{1}(p)\, +\, 
V_{2}(q) \sum_{n=-\infty}^{\infty} \delta(t-n), \] the quantum map 
is 
\begin{equation} 
U_{n, n^{\prime}}= \frac{1}{N} \exp(-2 \pi i N
V_{2}(\frac{n+a}{N})) \sum_{m=0}^{N-1} \exp(-2 \pi i N V_{1}(\frac{m+b}{N}))
\exp(\frac{2 \pi i}{N} (m+b)(n-n^{\prime}))
\end{equation} 
For the Harper map used in this paper we have taken $(a=b=1/2)$,
$V_{1}(p)=-g_{1}\cos(2 \pi p)/(2 \pi)$ and $V_{2}(q)=-g_{2}\cos(2 \pi
q)/(2 \pi)$. This set once again ensures symmetry properties of the
quantum map. The classical map is $( q_{t+1}=q_{t}+g_{1} \sin(2
\pi p_{t+1}), \; \, p_{t+1}= p_{t}-g_{2} \sin( 2 \pi q_{t}))$,
again the mod 1 rule is assumed.
If $g_{1,2}$ are equal, as assumed in this paper, the transition
to chaos occurs around $g_{1,2}=0.63$. In both the kicked Harper
model and the rotor the time between kicks has been taken as unity, as
anyway there are two parameters in the quantum problem, the scaled
Planck constant $N$ and the kick strength ($K$ or $g$).

The bakers map \cite{BVbak,Sarabak} comes in several varieties. 
It has been shown that the
usual bakers map $((q_{t+1}=2 q_{t}, \; p_{t+1}=p_{t}/2)\; \;  \mbox{if}
\; \;  0<q_{t}<1/2 \; \; \mbox{and} \; \; (q_{t+1}=2 q_{t}-1, \; p_{t+1}=(p_{t}+1)/2)
\; \; \mbox{if} \; \; 1/2<q_{t}<1 )$ is isomorphic to the $(1/2,1/2)$
Bernoulli process. The quantum bakers map on the torus is then the
unitary operator
\begin{equation}
U=G_{N}^{-1} \,  \left( \begin{array}{cc}
                        G_{N/2}&0\\
                        0&G_{N/2}
                        \end{array}
                        \right),
\end{equation}
where $G_{N}$ is the finite Fourier transform matrix with elements
\[
(G_{N})_{m,n}=\frac{1}{\sqrt{N}} \exp(-2 \pi i (m+1/2)(n+1/2)/N).
\]
We have assumed anti-periodic boundary conditions $(a=b=1/2)$ as this is known 
to preserve classical symmetries. 

The generalized bakers map used in this paper is a dynamical system 
implementing the $(2/3,1/3)$ Bernoulli process and is the classical map
$((q_{t+1}=3 q_{t}/2, \; p_{t+1}=2p_{t}/3) \; \;   \mbox{if} \; \; 
0<q_{t}<2/3 \; \; 
\mbox{and}  \; \; (q_{t+1}=3 q_{t}-2, \; p_{t+1}=(p_{t}+2)/3) \; \;  \mbox{if}
 \; \;   
2/3<q_{t}<1  )$. The quantum map is the unitary operator 
\begin{equation}
U=G_{N}^{-1} \,  \left( \begin{array}{cc}
                        G_{2N/3}&0\\
                        0&G_{N/3}
                        \end{array}
                        \right).
\end{equation}

\newpage

\begin{figure}[h]
\hspace*{.5in}\psfig{figure=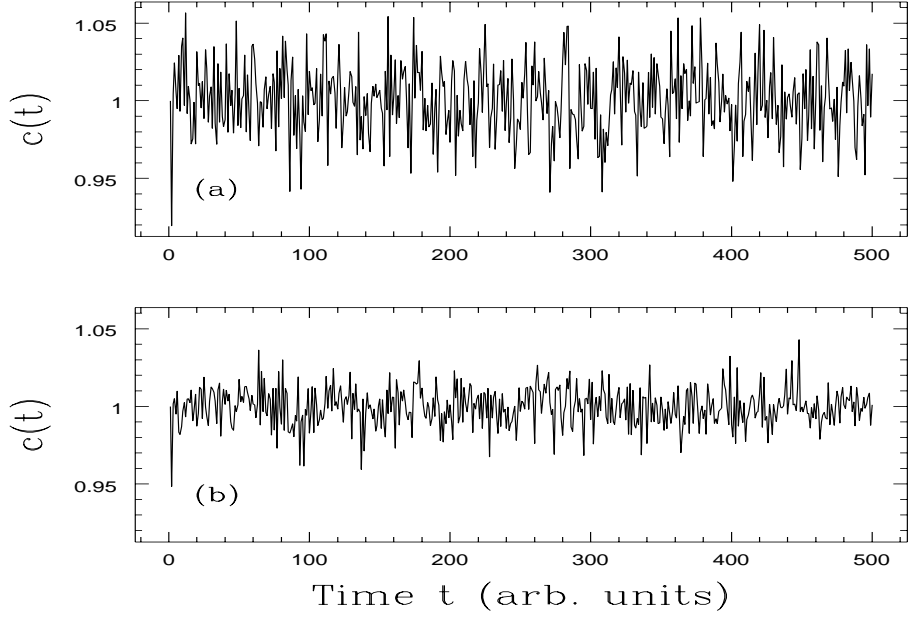,height=3.5in,width=5in}
\vspace{0.1in}
\caption{The relaxation fluctuations as a function of time for the
standard map with $K=20$ and (a) $N=100$ (b) $N=200$.}
\label{fig:1}
\end{figure}

\begin{figure}[h]
\hspace*{.5in}\psfig{figure=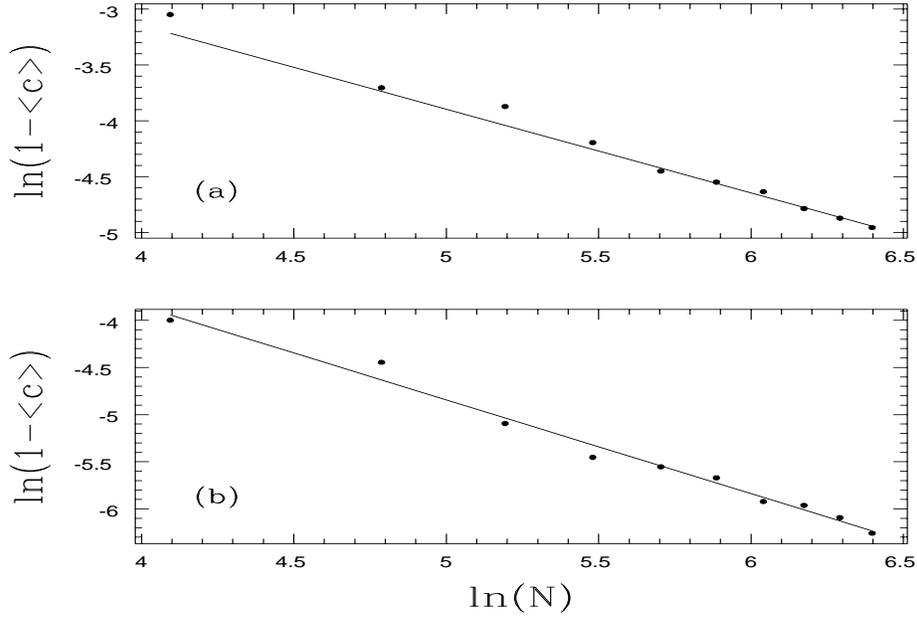,height=3.5in,width=5in}
\vspace{0.4cm}
\caption{ The average of the fluctuations for (a) the quantum bakers
map quantizing the (2/3,1/3) Bernoulli scheme and (b) the standard map with $K=20$. Shown 
is the logarithm of the
deviation from unity  as a function of $\log(N)$.}
\label{fig:2}
\end{figure}

\begin{figure}[h]
\hspace*{.5in}\psfig{figure=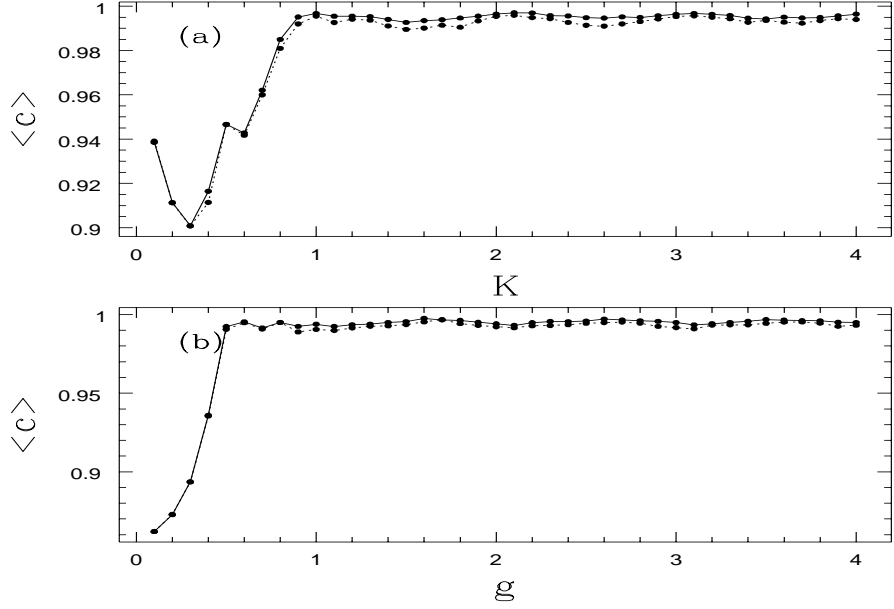,height=3.5in,width=5in}
\vspace{0.4cm}
\caption{The average relaxation as a function of the kicking strength
for (a) the standard map, (b) the Harper map. The solid line corrsponds to $N=300$
while the
dotted one to $N=200$.}
\label{fig:3}
\end{figure}

\begin{figure}[h]
\hspace*{.5in}\psfig{figure=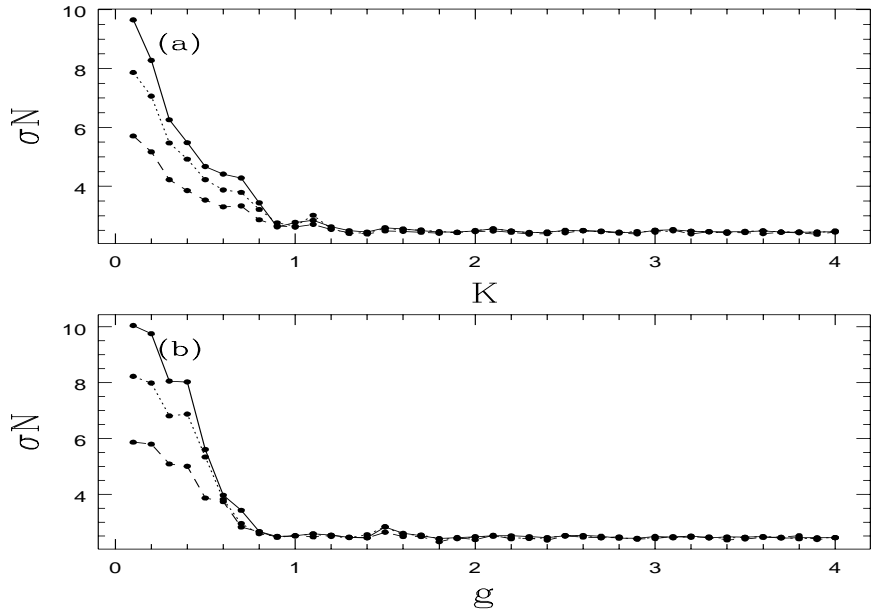,height=3.5in,width=5in}
\vspace{0.4cm}
\caption{The scaled standard deviation $\sigma N$ as a function of the
kick strength for (a) the standard map, (b) the Harper map. The solid line corresponds
to $N=300$, the dotted line to N=200 and the dashed line to N=100.}
\label{fig:4}
\end{figure}

\begin{figure}[h]
\hspace*{.5in}\psfig{figure=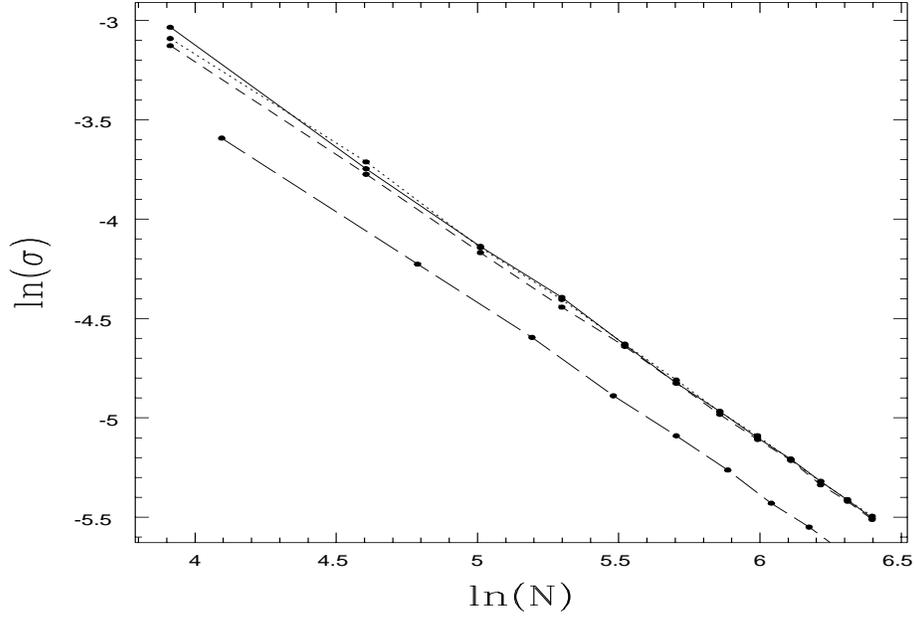,height=3.5in,width=5in}
\vspace{0.4cm}
\caption{The standard deviation as function of the scaled inverse
Planck constant $N$ on a logarithmic plot. The solid line corresponds to a standard
map with $K=20$, the dotted line to the Harper map with $g=4$, the small dashed line
to the usual symmetric quantum bakers map and the large dashed line to the unsymmetric
$(2/3,1/3)$ bakers map.}
\label{fig:5}
\end{figure}

\begin{figure}[h]
\hspace*{.5in}\psfig{figure=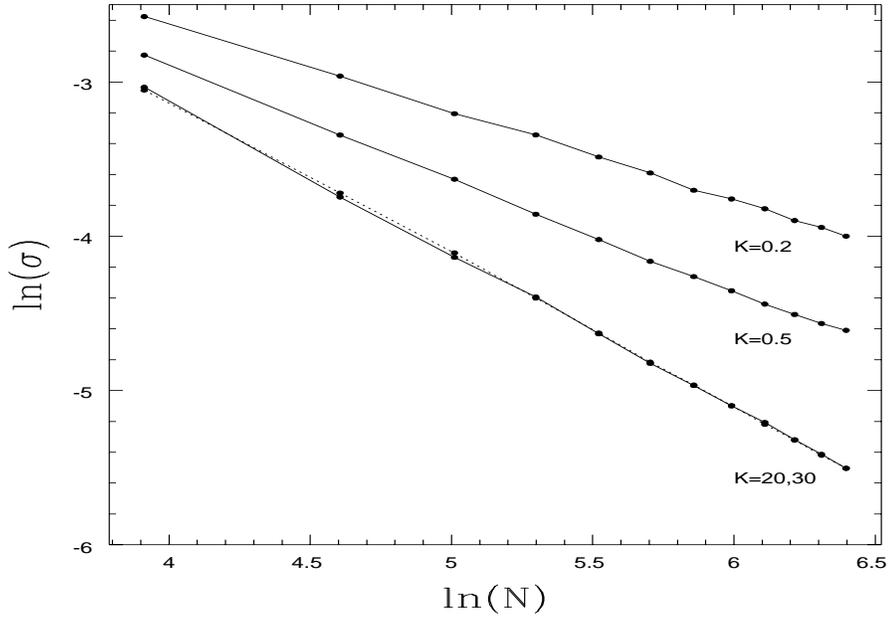,height=3.5in,width=5in}
\vspace{0.4cm}
\caption{Same as Fig. 5 for the standard map at different values of the
kick strength $K$, the dashed line corresponds to $K=30$ while the solid to $K=20$.
}
\label{fig:6}
\end{figure}

\begin{figure}[h]
\hspace*{.5in}\psfig{figure=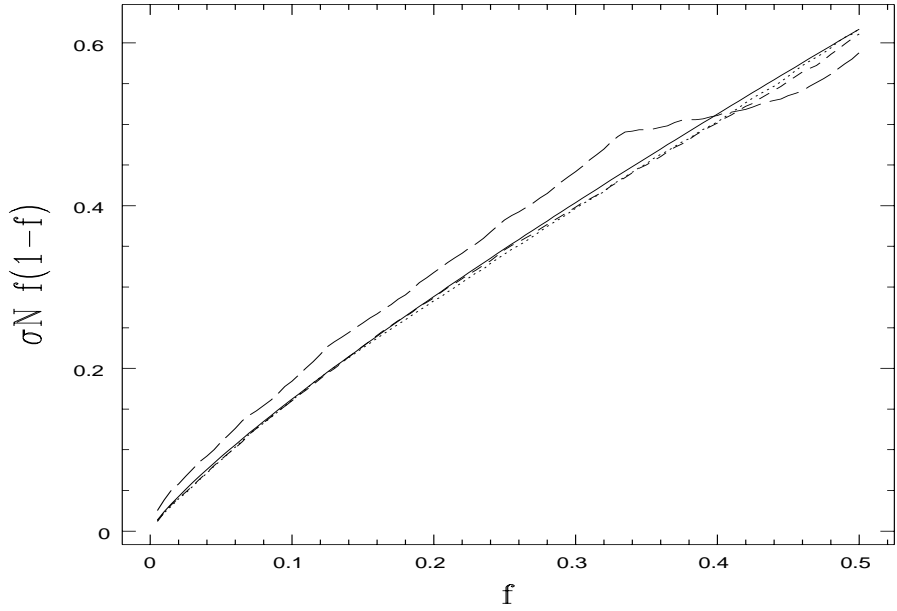,height=3.5in,width=5in}
\vspace{0.4cm}
\caption{The standard deviation normalized to $\sigma N f(1-f)$
as a function of $f$. The solid line is the fit (see text), while the dotted line corresponds to 
the standard map at $K=20$, the short dashed line to
the Harper map with $g=4$ and the long dashed line to the usual bakers map.
In all cases $N=200$.}
\label{fig:7}
\end{figure}

\begin{figure}[h]
\hspace*{.5in}\psfig{figure=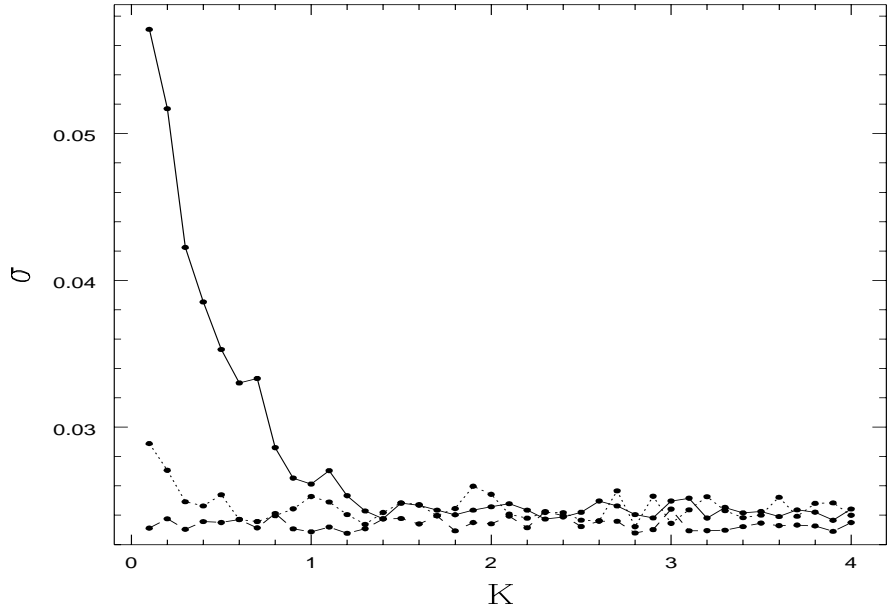,height=3.5in,width=5in}
\vspace{0.4cm}
\caption{The standard deviation as a function of the kick strength
for the standard map with $N=200$. The dotted line corresponds to using projection
operators constructed out of every alternative position basis and the dashed line to
projection operators constructed out of the eigenbasis of the standard map at $K=155$.
For comparison is shown the case of the usual projection operator from Fig. 4.}
\label{fig:8}
\end{figure}

\begin{figure}[h]
\hspace*{.5in}\psfig{figure=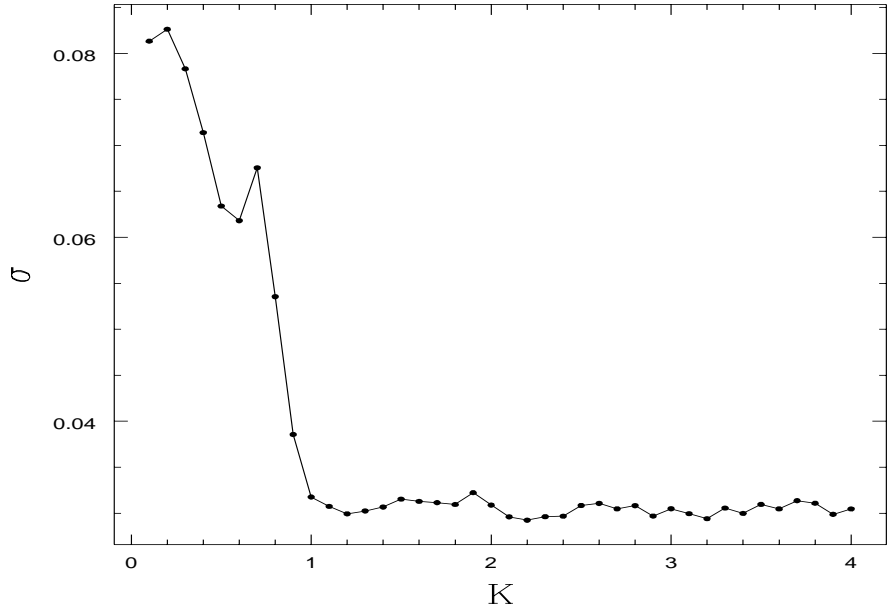,height=3.5in,width=5in}
\vspace{0.4cm}
\caption{The standard deviation as a function of kick strength
for the standard map using the operators constructed out of toral coherent states,
see text.}
\label{fig:9}
\end{figure}

\begin{figure}[h]
\hspace*{.5in}\psfig{figure=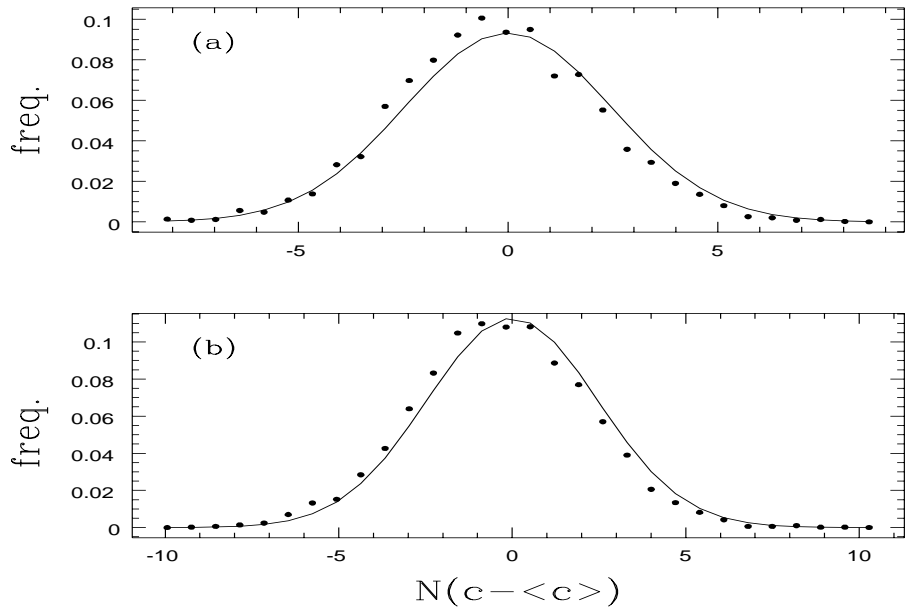,height=3.5in,width=5in}
\vspace{0.4cm}
\caption{The distribution of the relaxation fluctuations using
the first 5000 $c(t)$ values for the standard map with $K=20$ and (a) N=100 (b) N=200.
The solid
line corresponds to a Gaussian distribution.}
\label{fig:10}
\end{figure}

\begin{figure}[h]
\hspace*{.5in}\psfig{figure=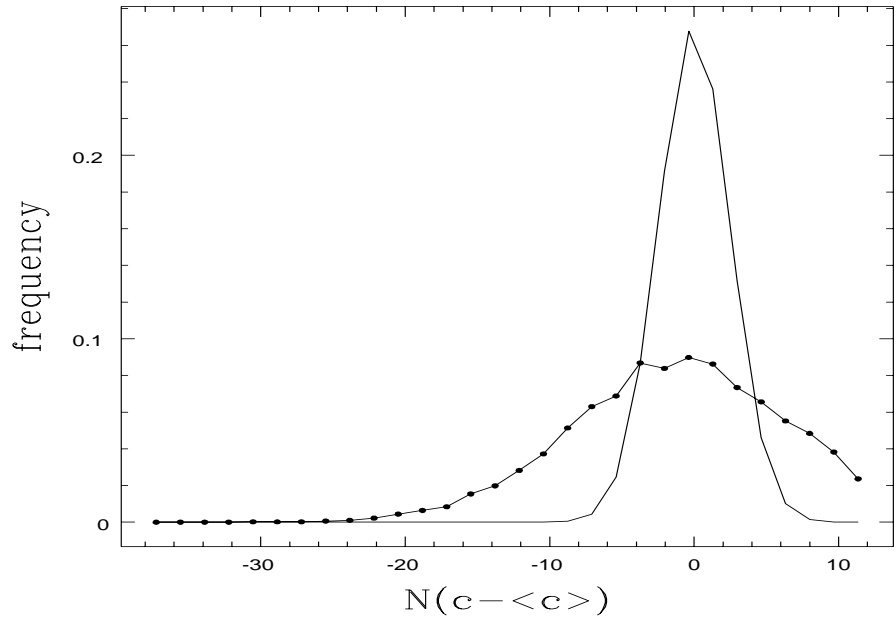,height=3.5in,width=5in}
\vspace{0.4cm}
\caption{Same as Fig. 10 but for the case when $K=0.2$ corresponding
to predominantly regular motion. The solid line still corresponds to the same Gaussian.}
\label{fig:11}
\end{figure}
 
\end{document}